\documentclass[12pt]{article}
\usepackage{amsmath,latexsym}
\usepackage{graphicx}
\begin{document}

\setlength{\unitlength}{1mm}

 \title{Existence of Wormholes in Einstein-Kalb-Ramond space time }
 \author{\Large $F.Rahaman^*$ , $M. Kalam^{**}$,  and $ A. Ghosh^* $ }
\date{}
 \maketitle
 \begin{abstract}
                  In recent, Kar.S et.al [ Phys Rev D 67,044005
                  (2003) ] have obtained static spherically
                  symmetric solutions of the Einstein-Kalb-Ramond
                  field equations. We have shown that their
                  solutions, indeed, represent Wormholes.

  \end{abstract}



 \bigskip
 \medskip
  \footnotetext{ Pacs Nos :  04.70 Dy,04.20 Gz,04.50 + h   

                              $*$Dept.of Mathematics, Jadavpur University, Kolkata-700 032, India\\
                                    E-Mail:farook\_rahaman@yahoo.com

                             $**$Dept. of Phys. , Netaji Nagar College for Women ,
                                          Regent Estate,
                                          Kolkata-700092, India
                              }

    \mbox{} \hspace{.2in}
In Einstein-Cartan theory, the symmetric Christoffel connection
is modified with the introduction of an anti-symmetric tensorial
term,known as space time torsion, which is presumed to have a
direct relation with spin [1]. It has been shown that the
massless anti-symmetric tensor Kalb-Ramond field, $B_{\mu\nu}$
equivalent to torsion is an inherent feature in the low energy
effective string action. Several authors [2] have shown that the
presence of Kalb-Ramond field in the background space time may
lead to various interesting astrophysical and cosmological
phenomena such as cosmic optical activity, neutrino lelicity flip,
parity violation etc. Recently, Kar et.al [3] have carried out
the most general study of the existence carried out of possible
spherical symmetric solutions of the Einstein-Kalb-Ramond field
equations. They have studied gravitational lensing and perihelion
precession in these space-time. They have also shown that for a
special case, one can get wormhole for a real Kalb-Ramond field.
In this article, we have shown that their general solutions,
indeed, always represent wormholes. According to the formation in
[3], the action is given by
\begin{equation}
                 S = \int d^4x \sqrt{-g}[\frac{R(g)}{k} -
                 \frac{1}{12} H_{\mu\nu\lambda}H^{\mu\nu\lambda}]
            \label{Eq1}
          \end{equation}
where $R(g)$ is Ricci scalar curvature and $H_{\mu\nu\lambda}$ is
Kalb-Ramond field strength \linebreak and $k = 8 \pi G $. \\

The field equations are
\begin{equation}
                  R_{\mu\nu} - \frac{1}{2} g_{\mu\nu} R = k
                  T_{\mu\nu}
            \label{Eq2}
          \end{equation}
\begin{equation}
                 D_\mu H^{\mu\nu\lambda} = \frac{1}{\sqrt{-g}}
                 \partial_\mu (\sqrt{-g}H ^{\mu\nu\lambda}) = 0
            \label{Eq3}
          \end{equation}
$T_{\mu\nu}$ is a symmetric two tensor which is analogous to the
energy momentum tensor and is given by
\begin{equation}
                 T_{\mu\nu}=\frac{1}{4}( 3 g_{\nu\rho}
                 H_{\alpha\beta\mu}H^{\alpha\beta\rho} -
                 \frac{1}{2}g_{\mu\nu}H_{\alpha\beta\gamma} H ^{\alpha\beta\gamma})
            \label{Eq4}
          \end{equation}
The asymmetric property of $ H_{\mu\nu\lambda}$ implies it has
only four independent components . According to Sengupta and Sur
[4] , the only nonzero component is $ H_{023}$. We denote $
H_{023}H^{023} = [h(r)]^2 $. \\

Kar et. al [3] have considered the general spherical symmetric
metric structure as
\begin{equation}
                ds^2 = B(r) dt^2 - \frac{dr^2}{A(r)}-r^2 d\Omega_2^2
            \label{Eq5}
          \end{equation}

From the gravitational field equations, they have obtained the
solutions as [3]
\begin{equation}
 B(r) =  1+  \frac{c_1}{r}+\frac{b c_1}{6r^3}-\frac{b
 c_1^2}{6r^4}+ \frac{6 b c_1^3 + 3 b^2 c_1}{40 r^5} + ..........
       \label{Eq6}
    \end{equation}
\begin{equation}
 A(r) =  1+  \frac{c_1}{r}-\frac{b }{r^2}+\frac{b
 c_1}{2r^3}- \frac{b c_1^2}{3 r^4} + \frac{1}{4 r^5}(bc_1^3 + \frac{b^2c_1}{6})..........
       \label{Eq7}
    \end{equation}
\begin{equation}
 h(r) =  \sqrt{\frac{b}{\bar{k}}} \frac{1}{r^2}[1-  \frac{c_1}{r}+\frac{c_1^2}{r^2}
 - \frac{1}{ r^3}(c_1^3 + \frac{bc_1}{6})..........]
       \label{Eq8}
    \end{equation}
where $ \bar{k} = \frac{3k}{4} $  and b, $c_1$ are arbitrary
constants.
\\
They have noticed that if $c_1 = 0 $ , then $ B(r) = 1 ,  A(r) = 1
- \frac{b}{r^2} $ which represent a wormhole.\\
Now, we shall show that general solutions (6) and (7) always
represent wormholes. To show this we rewrite the metric into the
Morris-Thorne Canonical form [5]
\begin{equation}
                ds^2 = e^{2f(r)} dt^2 - \frac{1}{[1 - \frac{b(r)}{r}]}dr^2-r^2 d\Omega_2^2
            \label{Eq9}
          \end{equation}
where,   $ r     \epsilon   (-\infty , +\infty) $. \\
To represent a wormhole , one must impose the following
conditions on the  metric (9) as [6] :  \\
1) The redshift function, $f(r)$ must be finite for all values of
$r$ . This means no horizon exists in the space time . \\
2) The shape function, $b(r)$ must obey the following conditions
at the throat $ r = r_0 $ : \linebreak $b(r_0) = r_0$ and
$b^\prime(r_0) < 1 $  [ these are known as Flair-out conditions ].\\
3) $\frac{b(r)}{r} < 1 $ for $ r >r_0 $ i.e. out of throat .\\
4)The space time is asymptotically flat i.e.  $\frac{b(r)}{r}
\rightarrow 0 $ as $ \mid r \mid \rightarrow \infty $  .\\

Here, the red-shift function and shape function take the form
\begin{equation}
                2f = \ln [ 1 +\frac{c_1}{r}+\frac{b c_1}{6
                r^3} - ........ ]
                \label{Eq10}
          \end{equation}
\begin{equation}
 b(r) =  -  c_1 +  \frac{b}{r}-\frac{bc_1}{2r^2}+ \frac{b
 c_1^2}{3r^3} - ..........
       \label{Eq11}
    \end{equation}
The throat of the wormhole occurs at $ r = r_0 $ where $ r_0 $
satisfies the equation $b(r) = r$. Suppose $\frac{1}{r} = y $,
then $b(r) = r$ implies
\begin{equation} g(y) = -c_1y + by^2 -
                \frac{bc_1}{2} y^3+ \frac{bc_1^2}{3} y^4  -  ............ - 1 = 0
            \label{Eq17}  \end{equation}
 Since $\frac{1}{r_0} $ is a root of equation
(12), then by standard theorem of algebra, either $g(y) > 0$ for $
y > \frac{1}{r_0}$ and $g(y) < 0 $ for $ y < \frac{1}{r_0}$ or
$g(y) < 0$ for $ y > \frac{1}{r_0}$ and $g(y) > 0$ for $ y <
\frac{1}{r_0}$. Let us take the first possibility and one can
note that for $ y = \frac{1}{r} < \frac{1}{r_0}$ i.e. $r>r_0$,
$g(y) < 0$, in other words,  $b(r) < r $. But when $ y =
\frac{1}{r} > \frac{1}{r_0}$ i.e. $r<r_0$, $g(y) > 0$, this means,
$b(r)
> r $, which violates the wormhole structure given in equation(9).

Here $ e^{2f(r)}\equiv B(r) $ has no zero for $ r \geq r_0 > 0 $
( as one can not assume $ r < r_0  $) because (1) at $ r
\rightarrow \infty $, $ B \rightarrow 1 $, so $ e^{2f(r)}\neq 0 $
at $ r \rightarrow \infty $ (2) if $ B(r) = 0  $ at $ r = r_0  $,
then $ B(r_0) - A(r_0) = 0 $ i.e.

\begin{equation}
  \frac{b}{r_0^2}-\frac{bc_1}{3r_0^3} + \frac{b
 c_1^2}{6r_0^4} + .......... = 0
       \label{Eq12}
    \end{equation}
But, $ A(r_0) \equiv [1 - \frac{b(r_0)}{r_0}] = 0 $ implies
\begin{equation}
   1+  \frac{c_1}{r_0}-\frac{b }{r_0^2}+\frac{b
 c_1}{2r_0^3}- \frac{b c_1^2}{3 r_0^4} + \frac{1}{4 r_0^5}(bc_1^3 +
 \frac{b^2c_1}{6})..........= 0
       \label{Eq7}
    \end{equation}
One can note that the above two equations could not hold
simultaneously.

Hence $ B(r_0) \neq 0 $. In other words, no horizon exists in the
space time.
 Also, $ \frac{b(r)}{r} \rightarrow 0 $ as $ \mid r \mid
\rightarrow \infty $. Thus the space-time with the solutions (6)
and (7) describes static spherically symmetric wormholes. \\

\pagebreak

 Retaining a few terms, the shape of the wormhole takes
the form:
\begin{figure}[htbp]
    \centering
        \includegraphics[scale=.8]{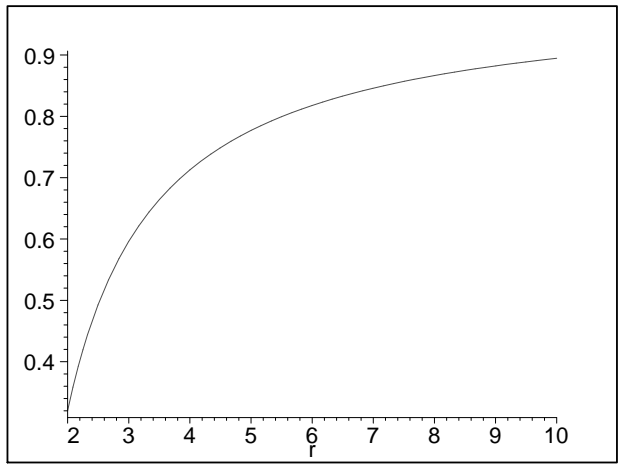}
        \caption{Shape of the wormhole for $c_1 = -1$ and $b = -1$}
   \label{fig:KRWH}
\end{figure}

 The asymptotical wormhole mass reads
\begin{equation}
                M  = \lim_{r\rightarrow \infty} \frac{1}{2}  b(r) =
                -
                \frac{1}{2} c_1
                \label{Eq13}
          \end{equation}

The axially symmetric embedded surface $ z = z(r)$ shaping the
Wormhole's spatial geometry is a solution of
\begin{equation}
                \frac{dz}{dr}=\pm
                \frac{1}{\sqrt{\frac{r}{b(r)}-1}}
            \label{Eq14}
          \end{equation}
By the definition of Wormhole, we can note that at the value $ r
= r_0 $ (the wormhole throat radius) equation (16) is divergent
i.e.
embedded surface is vertical there.\\
According to Morris and Thorne [5], the 'r' co-ordinate is
ill-behaved near the throat, but proper radial distance\\
\begin{equation}
 l(r) = \pm \int_{r_0^+}^r \frac{dr}{\sqrt{1-\frac{b(r)}{r}}}
            \label{Eq15}
          \end{equation}
 must be well behaved everywhere i.e. we must require that $ l(r)
 $is finite throughout the space-time. \\

\pagebreak
 For this Model,
\begin{equation}
 l(r) = \pm \int_{r_0^+}^r
 \frac{dr}{\sqrt{1-\frac{1}{r}[-  c_1 +  \frac{b}{r}-\frac{bc_1}{2r^2}+ \frac{b
 c_1^2}{3r^3} - ..........]}}
            \label{Eq16}
          \end{equation}
Though it is not possible to get the explicit form of the
integral but one can see that the above integral is a convergent
integral i.e.
proper length should be finite. \\

To summarize, we have shown that the static spherically symmetric
solutions of the Kalb-Ramond field equations obtained
by Kar et.al always represent  Wormholes. \\
According to Morris-Thorne [5], to keep a wormhole open, the
stress energy tensor of matter violates the null energy
conditions. As a result, the energy density of matter may be seen
as negative by some observer. Since maximum Kalb-Ramond energy
density is negative [3], it is clear that one can always
construct wormhole supported by Kalb-Ramond field. Finally, if we
take the parameter $c_1 < 0$, then asymptotic mass $'M'$ of the
Kalb-Ramond wormhole is positive i.e. a distant observer could
not see any difference of gravitational nature between Wormhole
and a compact mass 'M'.

        { \bf Acknowledgements }

          F.R is thankful to Jadavpur University and DST , Government of India for providing
          financial support under Potential Excellence and Young
          Scientist scheme .  \\


\end{document}